\documentclass[letter]{aa} 
\usepackage{amsmath,amssymb}
\usepackage{amstext}
\usepackage{natbib}
\usepackage{txfonts}
\usepackage{graphicx}
\usepackage{arydshln}
\bibpunct{(}{)}{;}{a}{}{,} 

\def\m2s2{\hbox{\,m$^{2}$\,s$^{-2}$}} 

\def\target{WD~1145+017}
 
\begin{document}

\title{Gray transits of WD 1145+017 over the visible band\thanks{Based on observations made with the Gran Telescopio Canarias (GTC), on the island of La Palma at the Spanish Observatorio Roque de los Muchachos of the IAC, and with the IAC80 telescope on the island of Tenerife at the Spanish Observatorio del Teide of the IAC, under Director's Discretionary Time.}}

\author{Alonso, R. \inst{1,2}
\and Rappaport, S. \inst{3}
\and Deeg, H.J. \inst{1,2}
\and Palle, E. \inst{1,2}
}

\institute{
Instituto de Astrof\'\i sica de Canarias, E-38205 La Laguna, Tenerife, Spain
\and Dpto. de Astrof\'isica, Universidad de La Laguna, 38206 La Laguna, Tenerife, Spain\label{La Laguna}
\and Department of Physics, and Kavli Institute for Astrophysics and Space Research, Massachusetts Institute of Technology, Cambridge, MA 02139, USA
}

\date{Received 14 March 2016 / Accepted 29 March 2016}
\abstract
{We have observed several relatively deep transits of the white dwarf WD 1145+017 with the Gran Telescopio Canarias (GTC) in the wavelength range 480 to 920 nm.  The observations covered approximately one hour on 2016 January 18 and two hours on 2016 January 20. There was variable extinction of the white dwarf during much of that time, but this extinction was punctuated by four sharp transits with depths ranging from 25\% to 40\%.  The spectrum was dispersed with a grism and the flux data were ultimately summed into four bands centered at 0.53, 0.62, 0.71, and 0.84 $\mu$m.  After careful normalization, we find that the flux light curves in all four bands are consistently the same, including through the deepest dips.  We use these results to compute \AA ngstr{\"om} exponents, $\alpha$, for the particles responsible for the extinction and find |$\langle \alpha \rangle $| $\lesssim 0.06$, assuming that the extinction is relatively optically thin. We use the complex indices of refraction for common minerals to set constraints on the median sizes of possible dust grains and find that particle sizes $\lesssim 0.5 \, \mu$m can be excluded for most common minerals.}


\titlerunning{WD~1145+017}
\authorrunning{Alonso et al.}

\maketitle

\section{Introduction}
\label{sec:intro}
The reported discovery of disintegrating planetesimals orbiting around the white dwarf \target\ \citep{Vanderburg:2015aa} adds to the growing evidence that a significant amount of white dwarfs (WDs) are accreting material orbiting within their tidal disruption radius. Over the last couple of decades, it has been established that around 25-50\% of WDs show metal lines in their spectra (\citealt{Zuckerman:2010aa,Koester:2014ab}) and at least 4\% are known to host dusty disks \citep{Rocchetto:2015aa}. \cite{Bear:2015aa} proposed a mechanism by which sub-Earth daughter planets at a few solar radii from the WD might form, requiring planetesimals with sizes of around 100~km. The putative disintegrating planetesimals orbiting around \target\ are thus excellent laboratories to test this and other theories. 

The Kepler K2 mission \citep{Howell:2014aa} light curve of \target\ showed transits of variable depth with several distinct periods in the range of 4.5 -- 5 h \citep{Vanderburg:2015aa}. Subsequent ground-based follow up  
(\citealt{Croll:2015aa,Gansicke:2016aa,Rappaport:2016aa})  revealed a complicated scenario in which eclipses with depths of up to 60\% were detected with durations ranging from $\sim$3 minutes to 12 minutes. The winter 2015 observing campaigns showed an increased activity in \target with more frequent and deeper eclipses.

Multiwavelength observations of transiting dust clouds have been used to constrain the particle sizes and compositions of material evaporating from the disintegrating planet candidates KIC12557548b (\citealt{Rappaport:2012aa, Croll:2014aa,Bochinski:2015aa}) and K2-22b \citep{Sanchis-Ojeda:2015aa}. Such constraints on the dust can be stronger when the depth of the transit is higher, owing to either larger clouds or smaller host stars. In the case of disintegrating planets, the hosts are main-sequence stars and the transiting clouds occult up to 1.5\% of the flux in the most favorable events. For \target,\, the very deep transits and their short duration, which helps to minimize the impact of instrumental systematics, allow for the most constraining determination to date of the hypothesized evaporating planetary material. Prior to the work presented here, two-band simultaneous photometry served to place 2-$\sigma$ limits on the average sizes of dust grains in WD 1145+017 of $\sim$0.15$\,\mu$m or larger, or $\sim$0.06$\,\mu$m or smaller \citep{Croll:2015aa}.    

In this Letter, we report the first high-cadence spectrophotometric observations of \target, with grism dispersed spectra from 0.5 to 0.9~$\mu$m on two different nights. The wide-band combined light curves show a complex flux variability, as reported in recent works (\citealt{Gansicke:2016aa,Rappaport:2016aa}), and a remarkable nonchromaticity that we use to place stronger constraints on the sizes and compositions of the occulting material. The constraints on the \AA ngstr{\"om} exponents are a factor $\sim$20 better than in previous studies of either WD 1145+017 or the known disintegrating planets.

\section{Observations and data analysis}
\label{sec:obs}
\begin{figure*}[ht]
\centering
\includegraphics[width=\textwidth]{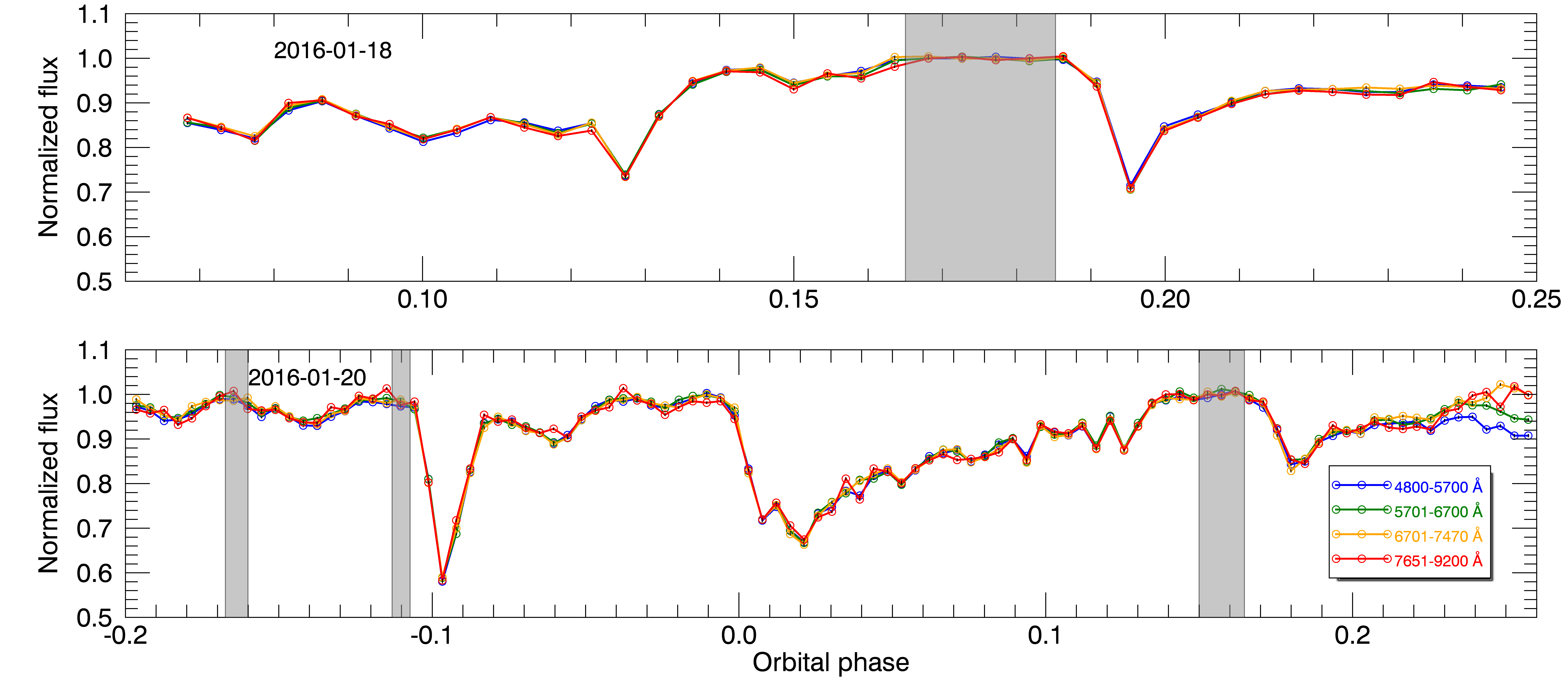}
\caption{Two extracted GTC light curves in the wavelength ranges indicated at the legend, showing several eclipse events that have comparable depths in all the different colors. The orbital phase is estimated using the ephemeris $\phi$=2457316.686+E*0.187517 from \cite{Rappaport:2016aa}. The points in the shaded regions were used for the normalization of the curves.} 
\label{fig:fig_lc}
\end{figure*}

We performed an intensive follow up of \target\ using the CAMELOT camera mounted at the IAC80 telescope during January 2016. The exposure times were fixed to 60~s, and no filter was used in order to collect a maximum number of photons. A deep eclipse ($\sim$40\% in flux) was observed on the night of 2016 January 16, and it was used to trigger observations at the 10.4~m Gran Telescopio Canarias (GTC) two nights later. 
We observed \target\ using the OSIRIS instrument \citep{Sanchez:2012aa} with the R300R grism, and we placed a nearby reference star (UCAC4 458-051099, at a distance of $\sim$1.9\arcmin) in the 12\arcsec-wide slit to perform differential spectrophotometry. The exposure time was fixed to 50~s, and the readout time was $\sim$24~s using the 200 kHz readout speed and a 2$\times$2 binning. 

Two sequences were obtained from 04:53 UT to 05:41 UT on the night of 2016 January 18, and a longer interval from 05:12 UT to 07:14 UT on the night of 2016 January 20. The last data points of the second sequence were taken during twilight, which occurred at 06:44 UT. The data were reduced using custom IDL routines. The images were calibrated using standard procedures, and the spectra of the target and the comparison star were extracted using a fixed aperture of 12 pixels centered on the spectral traces that were determined via Gaussian fits in the spatial direction. The zero order spectra, recorded in the images, was used to track the instrumental drifts in both spatial axes, and the wavelength calibration was updated accordingly. The flux from four different spectral regions covering the range from 4800\AA\ to 9200\AA\ was summed to produce four color light curves for the target and the reference star. To account for atmospheric effects, we computed the ratio between the target and the reference in all colors. The region between 7470\AA\ and 7651\AA\ was avoided as it contains a significant O$_2$ telluric absorption feature. 

After a first inspection of the light curves, showing a number of absorption events, different regions were selected to normalize each color curve. 
The final light curves built this way are represented in Fig.~\ref{fig:fig_lc}, and the shaded regions indicate the data points used for normalization. This normalization was performed by a simple median of the data points inside those regions. To provide an estimate of the precision of the data, we calculated the standard deviation of the ratio of each color curve with respect to the combined white light curve (thus not completely independent colors), resulting in 0.3 - 1\% for the four colors. The last data points on the second night show a significant deviation that we attribute to color effects due to the increased sky background at the end of the night, and the last six points were excluded from further analysis.

A cursory examination of the four-color light curves in Fig.~\ref{fig:fig_lc} shows a remarkable degree of agreement among the different bands.  We now proceed to interpret this achromaticity quantitatively in terms of dust scattering models.

\begin{figure}
\centering
\includegraphics[width=0.9\columnwidth]{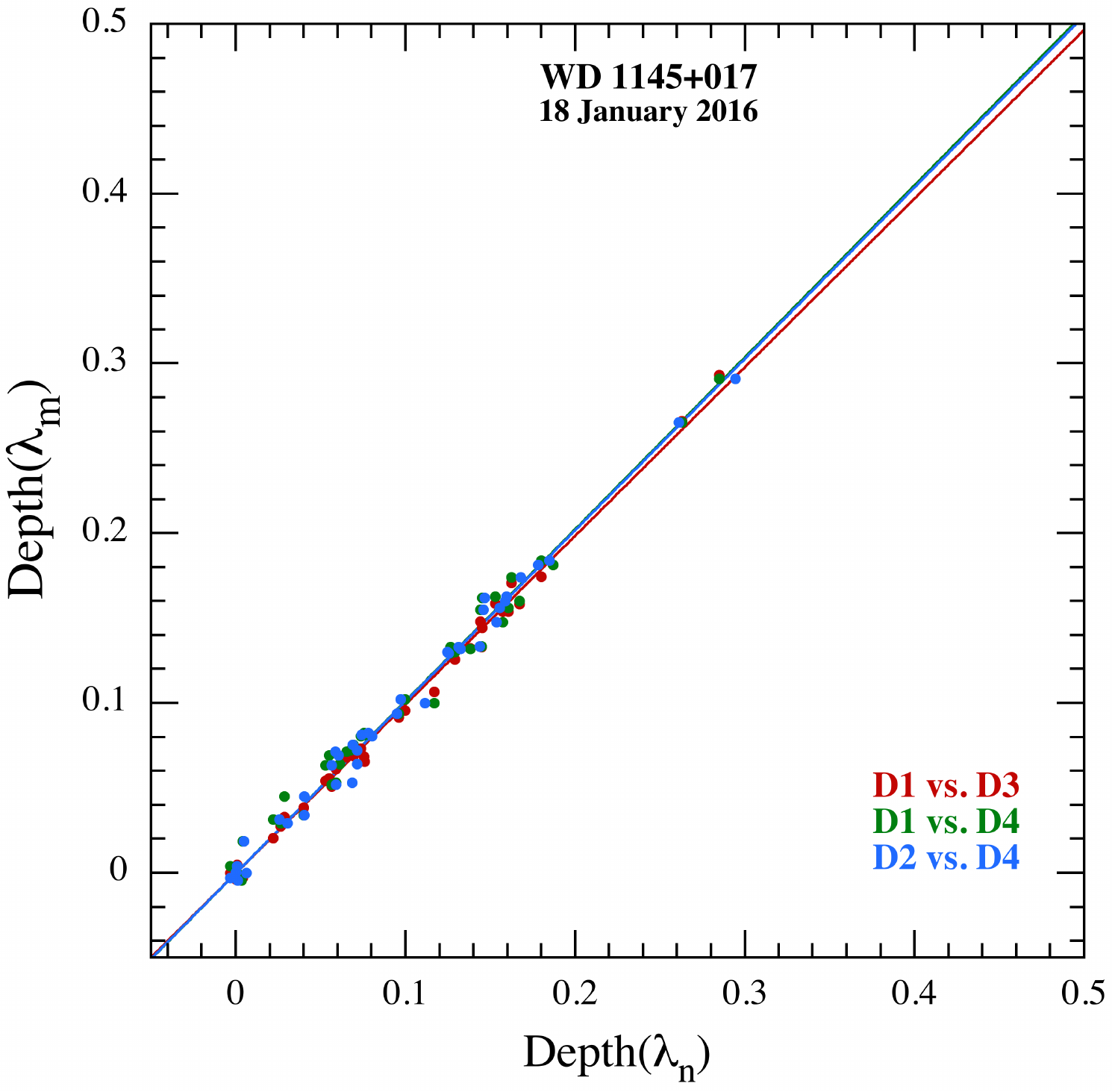} 
\includegraphics[width=0.9\columnwidth]{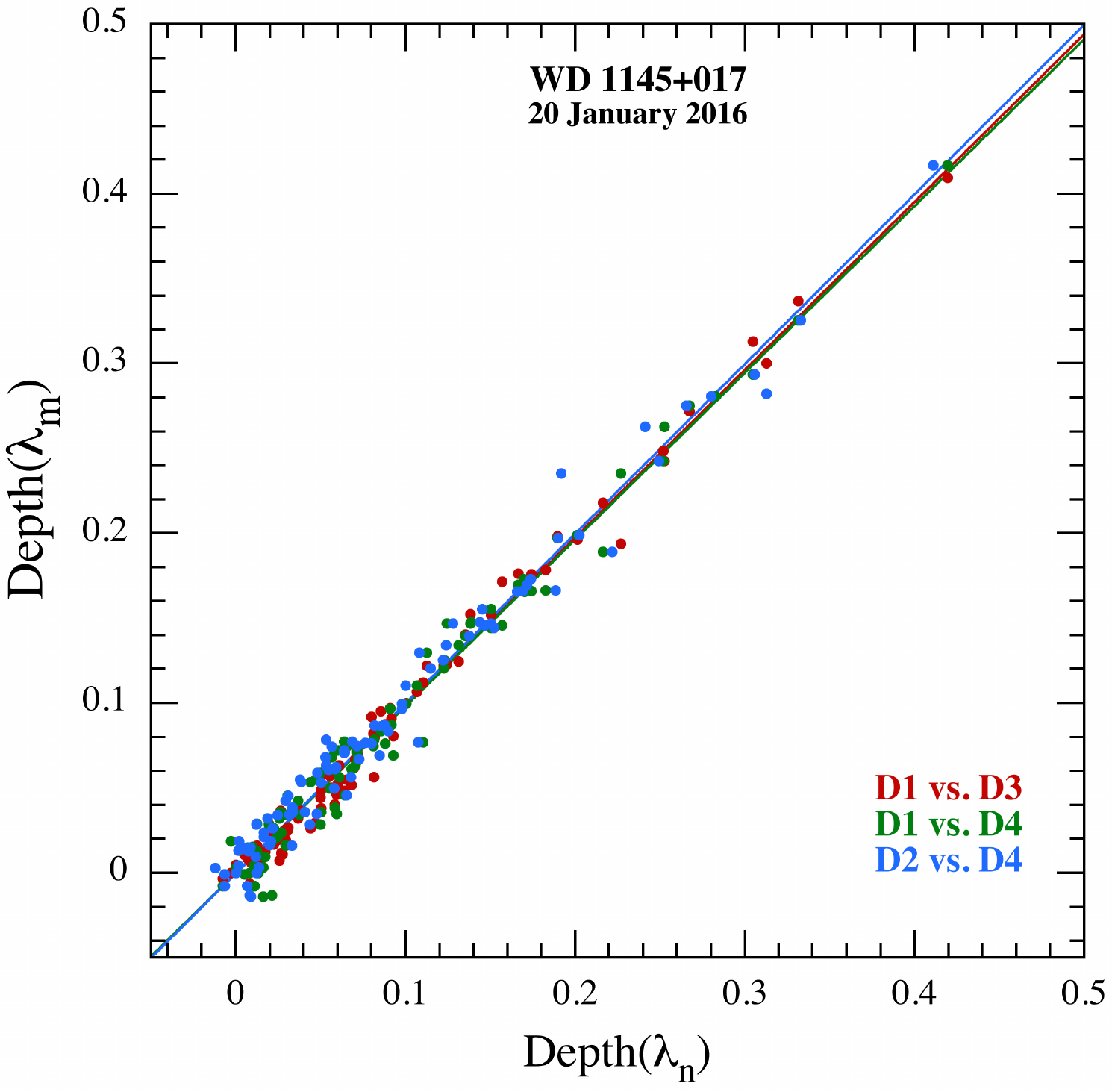}
\caption{Plots of the dip depths in a band centered on $\lambda_m$ vs.~the dip depths at $\lambda_n$ for three independent combinations of our four wavebands. The straight lines are $\chi^2$ fits to a function of the form $\lambda_{\rm m} = s_{\rm mn} \lambda_{\rm n}$. The upper and lower panels are for the data acquired on 2016 January 18 and 20, respectively.} 
\label{fig:ratios}
\end{figure}

\section{Assessing the \AA ngstr{\"o}m exponents}

If a cloud of small particles passes in front of a star, the fractional depth of the flux dip, $D$, that will be seen is somewhat involved, but can be written schematically as
\begin{equation}
D(\lambda_n) = \int_r \int_\phi \int_{\lambda_n-\Delta \lambda/2}^{\lambda_n+\Delta \lambda/2} S(r,\lambda)\left(1-e^{-\tau(r,\phi;\lambda)}\right)\,d\lambda \,d\phi \, r dr  
\label{eqn:dip1}
,\end{equation} 
where $\lambda_n$ is the mean wavelength over a bandpass $\Delta \lambda$; $S$ is the dimensionless surface brightness of the stellar disk at projected radial distance, $r$; $\phi$ is the azimuthal angle around the stellar disk; and $\tau$ is the extinction optical depth at any location over the stellar disk. The parameter  $S(r,\lambda)$ is taken to be normalized to unity over the stellar disk. For the purposes of this paper, we make three approximations, namely that: (i) $\tau$ is small; (ii) $\tau$ is constant over the fraction, $f$, of the star that the cloud covers; and (iii) the integral over the bandpass can simply be taken to be the evaluation at $\lambda_n$.  In that case, Eqn.~(\ref{eqn:dip1}) can be written much more simply as
\begin{equation}
D(\lambda_n) = \tau(\lambda_n) \,f ~.
\label{eqn:dip2}
\end{equation} 

\begin{table}
\centering
\caption{Summary wavelength dependence of dips}
\begin{tabular}{lcc}
\hline
\hline
Parameter\tablefootmark{a} &
2016 Jan.~18 &
2016 Jan.~20 \\
\hline
$s_{\rm 13}$ &  $0.994 \pm 0.007$ & $0.988 \pm 0.007$ \\ 
$s_{\rm 14}$ &  $1.013 \pm 0.011$  & $0.983 \pm 0.010$ \\
$s_{\rm 24}$ &  $1.010 \pm 0.009$ & $1.000 \pm 0.010$ \\ 
\hdashline
$\langle s_{\rm mn} \rangle$ & $1.006 \pm 0.010$ & $0.990 \pm 0.009$ \\
std.~dev.~$s_{\rm mn}$ & $0.010 \pm 0.006$ & $0.009 \pm 0.005$ \\
\hline
$\delta s_{\rm 13}$ &  $-0.006 \pm 0.007$ & $-0.012 \pm 0.007$ \\
$\delta s_{\rm 14}$ &  $+0.013 \pm 0.011$ & $-0.017 \pm 0.010$ \\ 
$\delta s_{\rm 24}$ &  $+0.010 \pm 0.009$ & $+0.000 \pm 0.010$ \\ 
\hdashline
$\langle \delta s_{\rm mn} \rangle$ & $0.006 \pm 0.010$ & $-0.010 \pm 0.009$ \\
std.~dev.~$\delta s_{\rm mn}$ & $0.010 \pm 0.006$ & $0.009 \pm 0.005$\\
\hline
$\ln(\lambda_3/\lambda_1)$ & 0.296 & 0.296 \\
$\ln(\lambda_4/\lambda_1)$ & 0.466 & 0.466 \\
$\ln(\lambda_4/\lambda_2)$ & 0.304 & 0.304 \\
\hdashline
$\langle \ln(\lambda_n/\lambda_m) \rangle$ & 0.354 & 0.354 \\
\hline
$\alpha_{\rm 13}$ &  $-0.020 \pm 0.024$ & $-0.040 \pm 0.024$ \\ 
$\alpha_{\rm 14}$ &  $+0.027 \pm 0.024$  & $-0.038 \pm 0.021$ \\
$\alpha_{\rm 24}$ &  $+0.033 \pm 0.030$ & $+0.001 \pm 0.033$ \\ 
\hline
\hline
$\langle \alpha \rangle$\tablefootmark{b}  & $-0.013 \pm 0.027$\tablefootmark{c} & $+0.026 \pm 0.032$\tablefootmark{c} \\
2-$\sigma$ limit on $|\alpha|$ & 0.06 & 0.06 \\\hline
\end{tabular}
\tablefoot{
\tablefoottext{a}{The value $s_{\rm mn}$ is the slope of the $D_{\rm m}$ vs.~$D_{\rm n}$ curve.  The uncertainties given are those that adjust $\chi^2_\nu$ to be equal to unity.  $\delta s_{\rm mn} \equiv s_{\rm mn} -1$.  The four centers of the wavebands are $\lambda_1 = 0.528$ nm; $\lambda_2 = 0.620$ nm; $\lambda_3 = 0.709$ nm; and $\lambda_4 = 0.840$ nm.  The values $\alpha_{\rm mn}$ are the \AA ngstr{\"o}m exponents computed for each of three combinations of wavebands, and  $\langle \alpha \rangle$ is the mean of the three values of \AA ngstr{\"o}m exponent computed by minimizing $\chi^2$.}  
\tablefoottext{b}{The uncertainties in $\langle \alpha \rangle$ are based on the rms scatter among the different results for $\alpha_{\rm mn}$ rather than on the formal statistical uncertainty.}
\tablefoottext{c}{If we also allow for an additive constant in Eqn.~(\ref{eqn:slope}) as a free parameter, i.e., to determine the flux = 1 level empirically, we find that these uncertainties would increase by $\sim$50\%.}
}
\label{tbl:ratios}
\end{table}

We now wish to quantify what the wavelength dependence of the dips tells us about the properties of the obscuring dust. Assuming that the extinction cross section and corresponding optical depth have a wavelength dependence of the form $\sigma_{\rm ext} \propto \tau \propto \lambda^{-\alpha}$, for a fixed dust column density and composition, this defines a quantity, $\alpha$, called the 
``\AA ngstr{\"o}m exponent''.  In the optically thin regime, and using Eqn.~(\ref{eqn:dip2}), we can write this as
\begin{equation}
\alpha_{\rm mn} \equiv ~ - \frac{\ln\left[D(\lambda_m)/D(\lambda_n)\right]}{\ln(\lambda_m/\lambda_n)}
.\end{equation}

In order to evaluate the ratios of $D(\lambda_m)/D(\lambda_n)$, we plot $D(\lambda_m)$ vs.~$D(\lambda_n$) in Fig.~\ref{fig:ratios} for three independent combinations of $n$ and $m$. We then fit these curves with a function of the form
\begin{equation}
D(\lambda_m) = s_{\rm mn} \, D(\lambda_n)
\label{eqn:slope}
.\end{equation}
The best-fit values of these slopes are reported in Table \ref{tbl:ratios}.  Because all the values of $s_{\rm mn}$ are within a few percent of unity, this immediately implies that the dips are at least approximately wavelength independent.  We therefore find it convenient to define a quantity $\delta s_{\rm mn}$ such that $s_{\rm mn} = 1+\delta s_{\rm mn}$.  In turn, this allows us to write a simple expression for the \AA ngstr{\"o}m exponent as
\begin{equation}
\alpha_{\rm mn} \simeq - \frac{\delta s_{\rm mn}}{\ln(\lambda_m/\lambda_n)}
.\end{equation}
These quantities are summarized in Table \ref{tbl:ratios}.

As we can see from Table \ref{tbl:ratios} the values of the \AA ngstr{\"o}m exponents are all rather close to zero, where  two values from the night of 2016 January 20 are only somewhat  significantly different from zero.  The weighted, in the sense of minimizing $\chi^2$, mean value of the \AA ngstr{\"o}m exponent for each of the nights is also very consistent with zero. We take the 1-$\sigma$ uncertainty on the average value of the \AA ngstr{\"o}m exponent to be the rms fluctuations of the three individual measurements of $\alpha_{mn}$ for each of the two observing nights.

From this analysis of $\langle \alpha \rangle$ we conclude that we have not detected any significant wavelength dependence of the dips and that a safe constraint is  $| \langle \alpha \rangle | \lesssim 0.06$ (at the 2-$\sigma$ confidence level).  For the purpose of comparing this to model calculations of the \AA ngstr{\"o}m exponents for different minerals, we take the effective wavelengths to be 0.574 nm and 0.774 nm.

\section{Discussion}
\label{sec:disc}

\begin{figure}
\centering
\includegraphics[width=0.9\columnwidth]{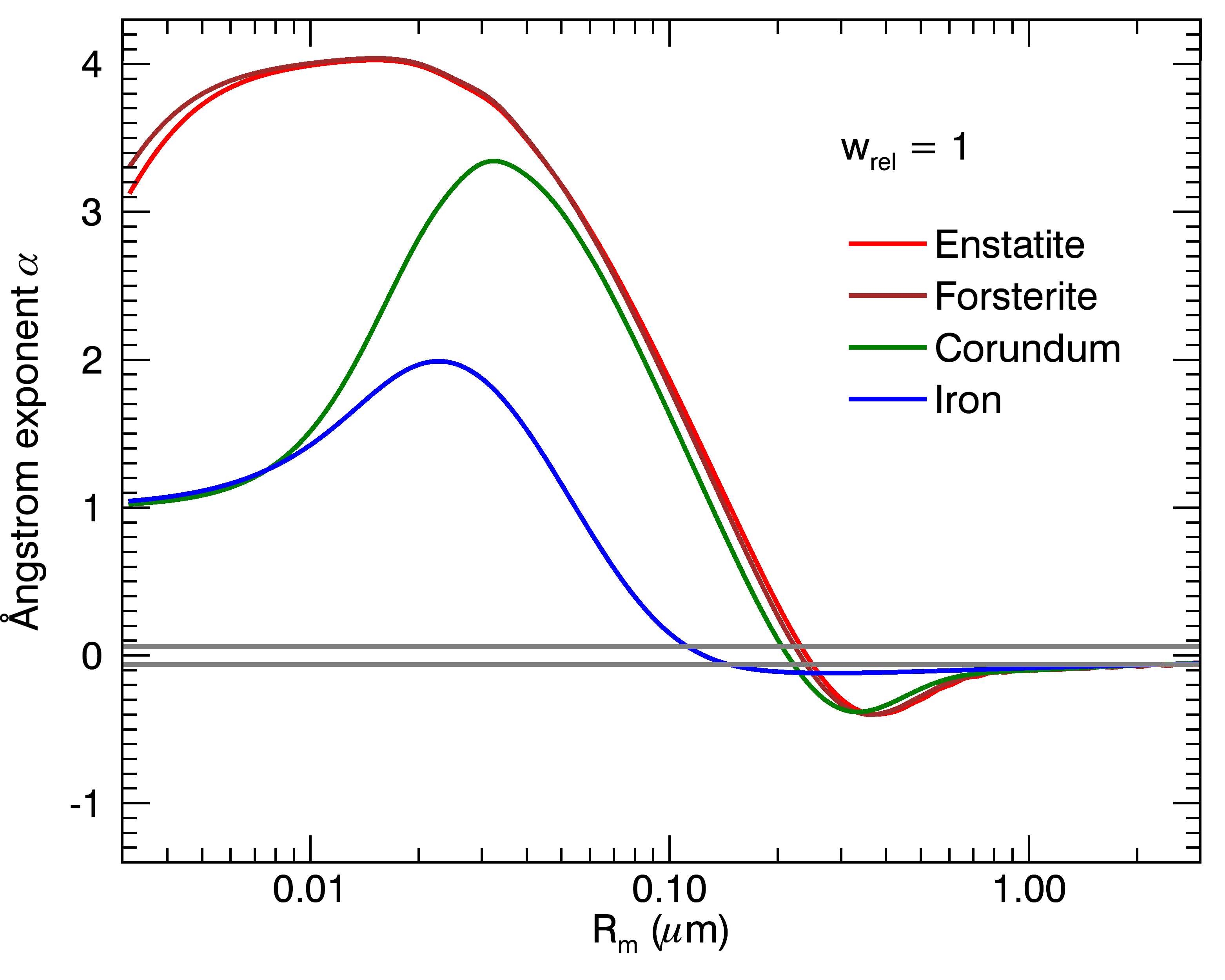} 
\includegraphics[width=0.9\columnwidth]{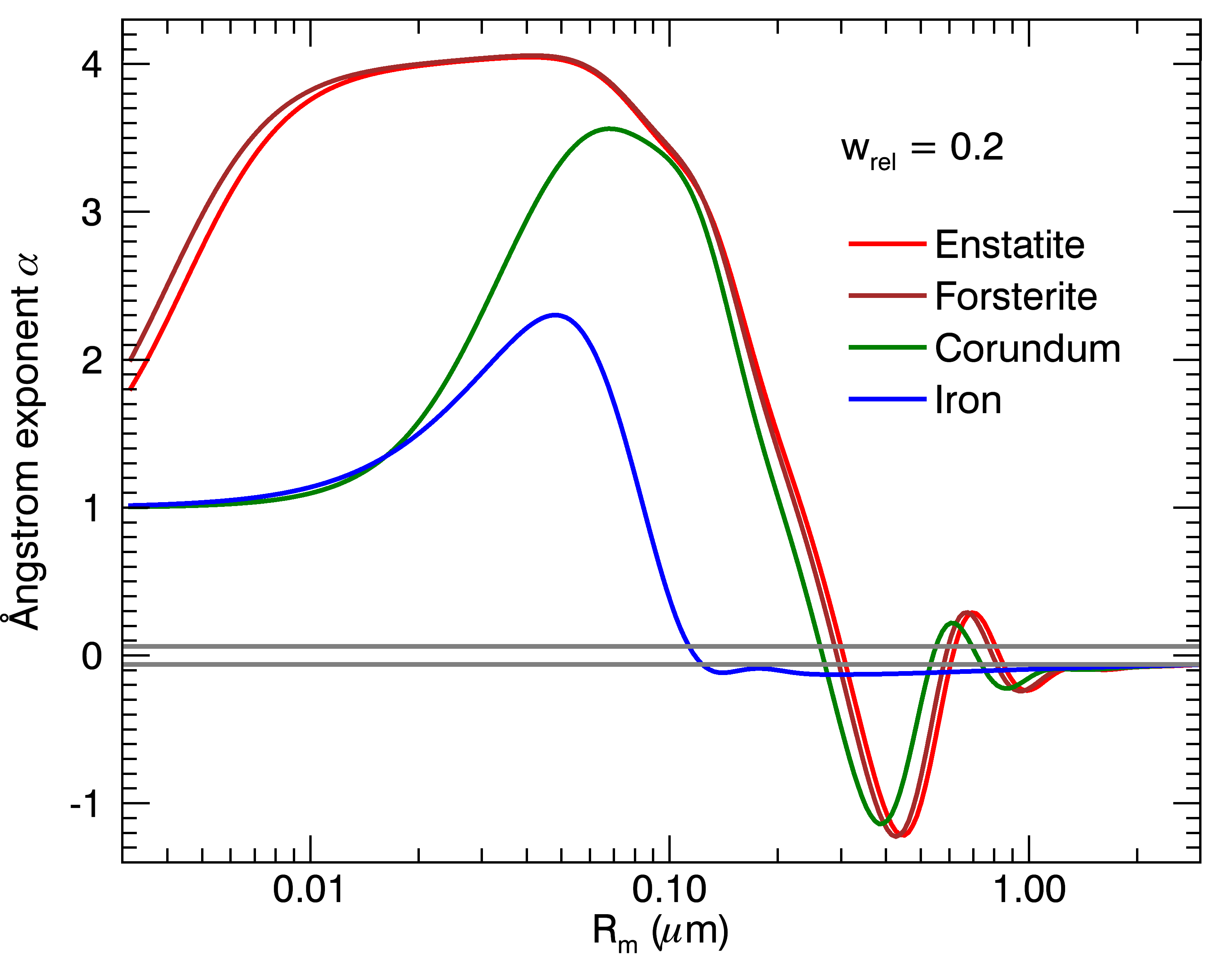}
\caption{Calculated \AA ngstr{\"o}m exponents $\alpha$ as a function of particle size for four different minerals. For each mineral we used a lognormal particle size distribution (top panel) and a narrower size distribution of 1/5th that width (bottom panel), where $R_m$ is the median particle size in the distribution. The two gray horizontal lines denote the 2-$\sigma$ lower and upper limits to $\alpha$ obtained from the GTC observations.} 
\label{fig:angexp}
\end{figure}

Figure~\ref{fig:angexp} shows the expected \AA ngstr{\"o}m exponents for assumed lognormal particle size distributions computed for the abovementioned effective wavelengths. We considered dust composed of the following minerals: enstatite ($n=1.66$,\  $\kappa=2.5\times10^{-5}$), forsterite ($n=1.69$,\  $\kappa=2\times10^{-5}$),  corundum ($n=1.76$,\  $\kappa=0.015$), and iron ($n=2.93$,\  $\kappa=3.2$), where $n$ and $\kappa$ refer to the real and imaginary indices of refraction, for which we have used the average values over the wavelength ranges considered. These \AA ngstr{\"o}m exponents were calculated from Mie extinction cross sections for spherical particles of given sizes and refractive indices. 

We use these \AA ngstr{\"o}m exponent curves to set lower limits on the dust particle sizes of an optically thin occulter.  Fig. \ref{fig:angexp} shows how the observational limit of $| \langle \alpha \rangle | \lesssim 0.06$ constrains the median particle size, $R_m$, in the assumed lognormal particle size distributions of two different widths. For all but iron, in the broader of the two lognormal distributions that we considered (upper panel of Fig.~\ref{fig:angexp}), $R_m$ must generally be $\ga 1\ \mu$m, while for the more narrow size distribution considered (lower panel of Fig.~\ref{fig:angexp}), $R_m \gtrsim 0.5 \, \mu$m is inferred.  However, in both cases there is also the formal, but somewhat contrived possibility that $R_m$ is very near to $0.25 \,\mu$m. For iron, $R_m$ is only restricted to median sizes of $\ga 0.1\ \mu$m.  
We note that the difference in permissible particle sizes is mainly driven by the refractive index $n$, with iron having a much higher refractive index than the other minerals. 

It is important to note that these particle size limits apply only to  relatively optically thin occulters in which the approximation from Eqn.~(\ref{eqn:dip1}) to Eqn.~(\ref{eqn:dip2}) is valid. As can be seen from these equations, there are infinite varieties of fractional coverage, $f$, and optical depth, $\tau$, that can account for the dip profiles.  In the limit where $\tau$ ranges from small to large in time, but $f$ is a constant, we simulated the type of curves presented in Fig.~\ref{fig:ratios} and found that the curves always take the slopes that appropriately reflect the actual \AA ngstr{\"o}m exponent before they converge on the line with unit slope at large $\tau$. In fact, the depth ratios are typically, significantly different from unity until $\tau$ becomes $\gtrsim 2$ unless $\alpha \simeq 0$. In the opposite limit where $\tau \rightarrow \infty$ and $f$ varies with time, it is not clear that the long, smooth transitions of the dips observed in WD 1145+017 (see, e.g.,  bottom panel of Fig.~\ref{fig:fig_lc};  phases $0-0.2$ with an egress duration of $\sim$54 min) can be simply explained. Since the time for a small object to cross the disk of the white dwarf is $\sim$1 min, this egress time is some 54 orbit-crossing times.  An optically thick cloud would have to act essentially as a knife edge that is aligned to within 1$^\circ$ of the orbital direction.  This seems contrived to say the least.  We therefore tentatively conclude that the observed dips are caused by variations in $\tau$ from very low to only moderate values.

We note that the dust occulters in WD 1145+017, at least during the two nights of our observations, are completely inconsistent with the dust size distribution inferred for the interstellar medium.  The ISM has an effective wavelength-dependent cross section that varies roughly as $\lambda^{-1}$ over the visible range (\citealt{Cardelli:1989aa,Fitzpatrick:1999aa}).  That would correspond to an \AA ngstr{\"o}m exponent of $\alpha \approx 1,$ which is ruled out by our observations.

Finally, we point out that the type of multicolor observation we conducted here should be repeated a number of times.  It is possible that the grain size distribution of the dusty effluents from debris orbiting WD 1145+017 may change with time.  This seems especially plausible because of the highly variable nature of the dust activity and the dip depths (see, e.g., G\"ansicke et al.~2016; Rappaport et al.~2016). In addition to repeating these observations, it would also be highly desirable to extend the wavelength coverage to J and K bands, where there may be a better chance to detect the wavelength dependence of the dip depths for larger grain sizes.

\begin{acknowledgements}
We thank an anonymous referee for very constructive and helpful comments. We acknowledge the Observatorio del Teide staff, especially the `TOTs', and the GTC staff for performing the observations in queue mode. RA acknowledges support by the Ram\'on y Cajal fellowship RYC-2010-06519. HD acknowledges support by grant AYA2012-39346-C02-02 from the Spanish Secretary of State for R\&D\&i (MINECO). RA, EP acknowledge funding from MINECO grants ESP2013-48391-C4-2-R and ESP2014-57495-C2-1-R. 
\end{acknowledgements}

\bibliographystyle{aa}
\bibliography{references}

\Online

\begin{appendix}
\onecolumn
\begin{longtab}
\begin{longtable}{lcccc}       
\caption{Light curve data 2016 January 18.} \\            
\hline\hline   
HJD\_UTC - &  Flux$_1$ & Flux$_2$  & Flux$_3$ & Flux$_4$ \\ 
2,400,000 & (4800-5700\AA) & (5701-6700\AA) & (6701-7470\AA) & (7651-9200\AA) \\
\hline
\endfirsthead
\caption{continued.}\\
\hline\hline   
HJD\_UTC - &  Flux$_1$ & Flux$_2$  & Flux$_3$ & Flux$_4$ \\ 
2,400,000 & (4800-5700\AA) & (5701-6700\AA) & (6701-7470\AA) & (7651-9200\AA) \\
\hline
\endhead
\hline
\endfoot
 57406.706817 &   0.85504 &    0.85618 &    0.86684 &	 0.86642 \\
 57406.707667 &   0.83935 &    0.84446 &    0.84625 &	 0.84387 \\
 57406.708518 &   0.82020 &    0.81532 &    0.82528 &	 0.81592 \\
 57406.709368 &   0.88321 &    0.88852 &    0.89334 &	 0.89977 \\
 57406.710219 &   0.90385 &    0.90502 &    0.90820 &	 0.90604 \\
 57406.711069 &   0.87100 &    0.87537 &    0.87435 &	 0.87000 \\
 57406.711919 &   0.84286 &    0.84638 &    0.84595 &	 0.85242 \\
 57406.712770 &   0.81292 &    0.82188 &    0.81838 &	 0.81870 \\
 57406.713621 &   0.83276 &    0.84132 &    0.84166 &	 0.83986 \\
 57406.714471 &   0.86169 &    0.86762 &    0.86779 &	 0.86796 \\
 57406.715321 &   0.85608 &    0.85415 &    0.85185 &	 0.84497 \\
 57406.716172 &   0.83760 &    0.83229 &    0.82929 &	 0.82576 \\
 57406.717022 &   0.85478 &    0.85339 &    0.85575 &	 0.83796 \\
 57406.717872 &   0.73711 &    0.73910 &    0.73399 &	 0.73470 \\
 57406.718723 &   0.87359 &    0.87477 &    0.86906 &	 0.87043 \\
 57406.719573 &   0.94365 &    0.94112 &    0.94907 &	 0.94783 \\
 57406.720423 &   0.97382 &    0.96981 &    0.97240 &	 0.97088 \\
 57406.721273 &   0.97794 &    0.97421 &    0.97951 &	 0.96860 \\
 57406.722124 &   0.94523 &    0.93953 &    0.94424 &	 0.93056 \\
 57406.722974 &   0.95992 &    0.95961 &    0.96154 &	 0.96601 \\
 57406.723824 &   0.97142 &    0.95976 &    0.96713 &	 0.95507 \\
 57406.724674 &   0.99603 &    0.99521 &    1.00279 &	 0.98129 \\
 57406.725525 &   0.99967 &    1.00000 &    1.00421 &	 0.99975 \\
 57406.726375 &   1.00000 &    1.00361 &    0.99892 &	 1.00288 \\
 57406.727225 &   1.00335 &    0.99944 &    1.00000 &	 0.99590 \\
 57406.728075 &   0.99931 &    0.99378 &    0.99537 &	 1.00000 \\
 57406.728925 &   0.99681 &    0.99908 &    1.00241 &	 1.00451 \\
 57406.729775 &   0.94749 &    0.94366 &    0.94565 &	 0.93679 \\
 57406.730626 &   0.71506 &    0.70568 &    0.70671 &	 0.70890 \\
 57406.731476 &   0.84710 &    0.84069 &    0.84132 &	 0.83719 \\
 57406.732326 &   0.87370 &    0.86879 &    0.86805 &	 0.86692 \\
 57406.733177 &   0.90049 &    0.90289 &    0.90455 &	 0.89769 \\
 57406.734027 &   0.92642 &    0.91976 &    0.92682 &	 0.91934 \\
 57406.734877 &   0.93267 &    0.92839 &    0.92982 &	 0.92770 \\
 57406.735727 &   0.93069 &    0.93104 &    0.93105 &	 0.92463 \\
 57406.736577 &   0.92419 &    0.92624 &    0.93449 &	 0.91868 \\
 57406.737428 &   0.92457 &    0.92204 &    0.93151 &	 0.91765 \\
 57406.738278 &   0.94107 &    0.93151 &    0.93864 &	 0.94678 \\
 57406.739128 &   0.93882 &    0.92841 &    0.93570 &	 0.93571 \\
 57406.739978 &   0.93497 &    0.94146 &    0.93195 &	 0.92869 \\
\hline
\end{longtable}
\label{tab:tab_phot}    
\end{longtab}
 
\begin{longtab}
\begin{longtable}{lcccc}       
\caption{Light curve data 2016 January 20.} \\            
\hline\hline   
HJD\_UTC - &  Flux$_1$ & Flux$_2$  & Flux$_3$ & Flux$_4$ \\ 
2,400,000 & (4800-5700\AA) & (5701-6700\AA) & (6701-7470\AA) & (7651-9200\AA) \\
\hline
\endfirsthead
\caption{continued.}\\
\hline\hline   
HJD\_UTC - &  Flux$_1$ & Flux$_2$  & Flux$_3$ & Flux$_4$ \\ 
2,400,000 & (4800-5700\AA) & (5701-6700\AA) & (6701-7470\AA) & (7651-9200\AA) \\
\hline
\endhead
\hline
\endfoot
57408.719842 &   0.97244 &    0.97550 &    0.98893 &	0.96608 \\
57408.720692 &   0.96361 &    0.97086 &    0.96777 &	0.95758 \\
57408.721543 &   0.94077 &    0.95256 &    0.95392 &	0.96518 \\
57408.722394 &   0.94372 &    0.94715 &    0.94158 &	0.93208 \\
57408.723244 &   0.95614 &    0.96168 &    0.97351 &	0.94648 \\
57408.724094 &   0.97752 &    0.97837 &    0.98313 &	0.97383 \\
57408.724945 &   0.98935 &    0.99858 &    0.98724 &	0.99570 \\
57408.725795 &   0.98923 &    0.99342 &    0.98572 &	1.00777 \\
57408.726645 &   0.97451 &    0.98130 &    0.99258 &	0.96788 \\
57408.727496 &   0.94976 &    0.95918 &    0.96192 &	0.96410 \\
57408.728347 &   0.96903 &    0.96849 &    0.97331 &	0.96555 \\
57408.729197 &   0.94967 &    0.95000 &    0.95119 &	0.94672 \\
57408.730047 &   0.93041 &    0.94109 &    0.93804 &	0.93752 \\
57408.730898 &   0.92981 &    0.94679 &    0.93306 &	0.93642 \\
57408.731748 &   0.95016 &    0.95625 &    0.95583 &	0.97135 \\
57408.732599 &   0.96252 &    0.96829 &    0.96304 &	0.96578 \\
57408.733449 &   0.98386 &    0.98653 &    0.99103 &	0.99686 \\
57408.734299 &   0.98292 &    0.98822 &    0.98752 &	0.99064 \\
57408.735150 &   0.97870 &    0.99196 &    0.98055 &	1.01343 \\
57408.736000 &   0.97317 &    0.98369 &    0.98831 &	0.97613 \\
57408.736851 &   0.97090 &    0.96699 &    0.97426 &	0.98386 \\
57408.737701 &   0.81054 &    0.81006 &    0.80178 &	0.80266 \\
57408.738551 &   0.58058 &    0.58900 &    0.59055 &	0.58307 \\
57408.739402 &   0.68743 &    0.68741 &    0.69962 &	0.71777 \\
57408.740252 &   0.82591 &    0.83105 &    0.82393 &	0.83380 \\
57408.741102 &   0.93604 &    0.93475 &    0.92518 &	0.95429 \\
57408.741953 &   0.94414 &    0.94532 &    0.94989 &	0.93912 \\
57408.742804 &   0.93910 &    0.93210 &    0.93669 &	0.94358 \\
57408.743654 &   0.91904 &    0.92838 &    0.91804 &	0.92518 \\
57408.744504 &   0.91392 &    0.91478 &    0.91361 &	0.91348 \\
57408.745355 &   0.88978 &    0.89267 &    0.88797 &	0.92303 \\
57408.746205 &   0.90892 &    0.90223 &    0.90291 &	0.90314 \\
57408.747055 &   0.94486 &    0.94234 &    0.94289 &	0.95032 \\
57408.747906 &   0.97384 &    0.96682 &    0.96336 &	0.96401 \\
57408.748758 &   0.98757 &    0.98809 &    0.98398 &	0.97123 \\
57408.749608 &   0.98400 &    0.99135 &    0.98815 &	1.01401 \\
57408.750459 &   0.99108 &    0.99182 &    0.99416 &	0.98669 \\
57408.751309 &   0.97554 &    0.98368 &    0.98063 &	0.97887 \\
57408.752159 &   0.96985 &    0.96918 &    0.97510 &	0.95446 \\
57408.753009 &   0.98054 &    0.98784 &    0.97103 &	0.97153 \\
57408.753859 &   0.99000 &    0.99660 &    0.98989 &	0.98484 \\
57408.754710 &   1.00294 &    0.99814 &    1.00000 &	0.98150 \\
57408.755560 &   0.99320 &    0.99142 &    0.99173 &	0.98495 \\
57408.756410 &   0.95312 &    0.96225 &    0.97044 &	0.94509 \\
57408.757261 &   0.83363 &    0.82843 &    0.82350 &	0.83030 \\
57408.758111 &   0.71731 &    0.71979 &    0.71922 &	0.71913 \\
57408.758962 &   0.74793 &    0.75084 &    0.75158 &	0.75717 \\
57408.759812 &   0.69543 &    0.69418 &    0.68693 &	0.70615 \\
57408.760662 &   0.66856 &    0.66733 &    0.66323 &	0.67454 \\
57408.761513 &   0.73268 &    0.73454 &    0.72798 &	0.72449 \\
57408.762363 &   0.74753 &    0.75876 &    0.75737 &	0.73723 \\
57408.763214 &   0.78367 &    0.77847 &    0.78183 &	0.81110 \\
57408.764064 &   0.77311 &    0.80832 &    0.80598 &	0.76458 \\
57408.764915 &   0.81752 &    0.81140 &    0.82127 &	0.83374 \\
57408.765765 &   0.83031 &    0.82623 &    0.83369 &	0.82681 \\
57408.766615 &   0.79887 &    0.79758 &    0.80339 &	0.80087 \\
57408.767466 &   0.82989 &    0.83446 &    0.83133 &	0.83432 \\
57408.768316 &   0.86142 &    0.85618 &    0.85257 &	0.85240 \\
57408.769167 &   0.86867 &    0.87616 &    0.87549 &	0.86589 \\
57408.770017 &   0.87563 &    0.87203 &    0.87712 &	0.85304 \\
57408.770867 &   0.84965 &    0.84779 &    0.84787 &	0.85543 \\
57408.771718 &   0.86491 &    0.86267 &    0.85963 &	0.86017 \\
57408.772568 &   0.88743 &    0.89203 &    0.87791 &	0.87022 \\
57408.773418 &   0.89964 &    0.90198 &    0.89976 &	0.90034 \\
57408.774269 &   0.86197 &    0.84985 &    0.84737 &	0.85296 \\
57408.775119 &   0.92823 &    0.92748 &    0.92977 &	0.93308 \\
57408.775970 &   0.91454 &    0.91024 &    0.90471 &	0.91638 \\
57408.776820 &   0.90834 &    0.91252 &    0.90915 &	0.91273 \\
57408.777670 &   0.92836 &    0.93635 &    0.93424 &	0.92927 \\
57408.778521 &   0.87752 &    0.88526 &    0.87761 &	0.87933 \\
57408.779371 &   0.95032 &    0.95183 &    0.94290 &	0.94094 \\
57408.780222 &   0.87735 &    0.87803 &    0.87731 &	0.87457 \\
57408.781072 &   0.92891 &    0.93599 &    0.92764 &	0.92901 \\
57408.781922 &   0.97883 &    0.97934 &    0.97647 &	0.98144 \\
57408.782773 &   0.98687 &    0.98783 &    0.99199 &	0.99998 \\
57408.783623 &   0.99494 &    1.00658 &    0.98946 &	1.00080 \\
57408.784473 &   0.98741 &    0.99251 &    0.98814 &	0.98724 \\
57408.785324 &   0.99222 &    1.00000 &    1.00613 &	1.00000 \\
57408.786174 &   1.00000 &    1.01208 &    0.99521 &	0.99729 \\
57408.787025 &   1.00782 &    1.00665 &    1.00358 &	1.00790 \\
57408.787875 &   0.98892 &    0.99854 &    0.99220 &	0.98695 \\
57408.788725 &   0.97077 &    0.98019 &    0.98055 &	0.98368 \\
57408.789576 &   0.92027 &    0.92392 &    0.90806 &	0.92322 \\
57408.790426 &   0.84328 &    0.85339 &    0.82852 &	0.85409 \\
57408.791276 &   0.84979 &    0.85477 &    0.85569 &	0.84456 \\
57408.792127 &   0.89358 &    0.89991 &    0.89328 &	0.88955 \\
57408.792977 &   0.90713 &    0.91538 &    0.91957 &	0.93071 \\
57408.793827 &   0.91809 &    0.91831 &    0.92025 &	0.91322 \\
57408.794678 &   0.91195 &    0.92015 &    0.91254 &	0.92395 \\
57408.795528 &   0.93179 &    0.94155 &    0.94835 &	0.93892 \\
57408.796413 &   0.93472 &    0.94355 &    0.94511 &	0.92556 \\
57408.797263 &   0.93622 &    0.93157 &    0.95165 &	0.92275 \\
57408.798113 &   0.93991 &    0.93619 &    0.94751 &	0.92788 \\
57408.798964 &   0.91857 &    0.94679 &    0.94364 &	0.92168 \\
57408.799814 &   0.94160 &    0.96681 &    0.95962 &	0.96157 \\
57408.800664 &   0.94882 &    0.98129 &    0.98681 &	0.96787 \\
57408.801515 &   0.94988 &    0.97578 &    0.98128 &	0.99740 \\
57408.802373 &   0.92148 &    0.97521 &    0.99243 &	1.00543 \\
57408.803225 &   0.92958 &    0.96170 &    1.02236 &	0.97246 \\
57408.804076 &   0.90754 &    0.94642 &    1.01382 &	1.01820 \\
57408.804927 &   0.90766 &    0.94351 &    0.99937 &	0.99822 \\
\hline
\end{longtable}
\label{tab:tab_phot2}    
\end{longtab}

\end{appendix}

\end{document}